\documentclass[twocolumn,amsmath,amssymb]{revtex4}

\usepackage{graphicx}
\usepackage{dcolumn}
\usepackage{bm}

\begin{document}


\title{Terahertz pulse generation via optical rectification in photonic crystal microcavities}

\author{A. Di Falco}
\email{difalco@uniroma3.it}
\affiliation{NooEL - Nonlinear Optics and OptoElectronics Laboratory\\ National Institute for the Physics of the Matter, University Roma Tre\\
Via della Vasca Navale, 84 - 00146 - Rome, Italy }

\author{C. Conti}
\email{claudio.conti@phys.uniroma1.it}
\altaffiliation{
Present address: INFM CRS-SOFT, Universita' di Roma ``La Sapienza,'' P. A. Moro,2  - 00185 -
Roma, Italy}

\author{G. Assanto}
\email{assanto@uniroma3.it}
\affiliation{NooEL - Nonlinear Optics and OptoElectronics Laboratory\\ National Institute for the Physics of the Matter, University Roma Tre\\
Via della Vasca Navale, 84 - 00146 - Rome, Italy }

\begin{abstract} 
Using a 3D fully-vectorial nonlinear time-domain analysis we numerically investigate the generation of terahertz radiation 
by pumping a photonic crystal microcavity out of resonance. 
High quality factors and a quadratic susceptibility lead to few-cycle terahertz pulses via optical rectification. 
Material dispersion as well as linear and nonlinear anisotropy is fully accounted for.
\end{abstract}

\maketitle

\noindent In the last decades the increasing need of terahertz (THz) sources for biomedical applications
and for spectroscopy has dragged the researchers' attention 
toward new methods to generate radiation in this wavelength region. \cite{Darmo}
Several ideas have been implemented, from semiconductor-based fotoconductive antennas 
to optical rectification in crystals. \cite{Davies,Ding,Lee,Shi}
Resonant structures have been proposed in order to improve THz generation efficiency; \cite{DarmoEE} in this context
Photonic Crystals (PC) offer various degrees of freedom for device design and optimization,
both in terms of dispersive features and nanoscale integration. \cite{Othaka,John,Yablonovitch} 
PC consist of a high index-contrast periodic distribution in either one, two or three spatial dimensions: they
severely affect light propagation when its wavelength is comparable to their characteristic period and,
since electromagnetic radiation can be confined in extremely small volumes for long times, 
PC lend themselves to a new class of microcavities with engineerable Quality Factors (Q) as high as $10^{6}$. \cite{Ryu,Zhang}
Several authors have recently recognized the advantages encompassed by the use of both linear and 
nonlinear resonant structures and PC in the THz region. \cite{DeLaRue,Lu,Tani,Zudov}
In this Letter we numerically show that light modulation at terahertz frequencies can be achieved by resorting to a 
high-Q PC microcavity fed by an out-of-resonance pump. 
Exciting a long-lived PC cavity-mode with a single source slightly detuned from resonance, taking 
advantage of transient oscillations in a quadratic medium we obtain THz radiation via optical rectification. 
The dynamics behavior depends on both detuning and Q. \\
In order to simulate the generation process we developed a three dimensional parallel Finite Difference Time Domain (FDTD) code, 
able to account for second-order nonlinearities as well as material anisotropy and dispersion. 
To this extent Maxwell equations are coupled to an oscillating dipole with an anharmonic term: \cite{YarivBook}

\begin{equation}
\label{Maxwell}
\begin{array}{l}
\displaystyle \nabla\times{\bf E}=-\mu_0 \frac{\partial {\bf H}}{\partial t},~~ 
\displaystyle \nabla\times{\bf H}=\epsilon_0 \frac{\partial {\bf E}}{\partial t}+\frac{\partial {\bf P}}{\partial t}, \\
\\
\displaystyle \frac {\partial^2{\bf P}}{\partial{t^2}}+2\gamma_{0}\frac{\partial{{\bf P}}}{\partial{t}}+\omega_{R}^{2}{\bf P}+\underline{\underline{\bf D}}:{\bf PP} =\epsilon_{0}(\epsilon_{s}-1)\hat{\omega}_{R}^2 {\bf E}.
\end{array}
\end{equation}

The last equation of (\ref{Maxwell}) assigns the material properties: $\bf{P}$  is the polarization 
due to a single-pole Lorentz-dispersion centered in $\omega_0$, and 
the other parameters yield a refractive index $n\simeq3.5$ 
at $\lambda=1.55\mu$m, with static permittivity $\epsilon_{s}=11.7045$, loss coefficient $\gamma_{0}=3.8e8$, 
$\omega_{R}=1.1406e16$ and $\hat\omega_{R}=1.0995e16$ in MKS units.
The tensor \underline{\underline{\bf D}} provides the quadratic nonlinear response, 
and includes a $\pi/4$ rotation of the crystallographic axes (with respect to the spatial coordinates) in order to generate a second-order polarization from a linearly polarized input.
All linear and nonlinear properties, including dispersion, match those of AlGaAs. \cite{YarivBook} .
The microcavity consists of a void in a six-hole PC-wire ($w=450$ nm, $h=270$ nm and $l=4 \mu$m, see Fig. 1 inset) designed to yield 
single-mode TE-like propagation in both input and output waveguides. \cite{PCwireAPB} 
For $a=450$ nm and $d=0.35a$, a resonant state appears near 
the middle band-gap at $\lambda_{0}=1.447 \mu$m, with Q=373. 
Fig.1 graphs the oscillation spectrum, obtained by exciting the device with a 
single-cycle (broad-band) pulse and Fourier-transforming the field as probed inside the structure. 
The arrow indicates the off-resonance input wavelength.\\
To reveal the transient due to microcavity excitation, we launched (in the input waveguide) an 
``\emph{mnm}'' pulse, \cite{Ziolkowski} which ensures precise control of both leading and trailing edges of the pulse and 
its bandwidth, i.e. avoids spurious spectral artifacts. \\
To theoretically assess the numerical results, we also employed coupled mode theory in the time domain for 
the case of a pumped cavity with feeding waveguide. \cite{HausBook}
The decay rate $1/\tau=2/(\omega_0Q)=1/\tau_o+2/\tau_e$  includes both internal losses ($1/\tau_o$) and those ($1/\tau_e$) 
due to the external load, i.e. input and output waveguides. 
For a step-like excitation $s(t)=0$ 
for $t<0$ and $s(t)=s\exp[j\omega t]$ for $t>0$, the time evolution of the cavity mode 
$a(t)$ is given by (the input power is $P_{input}=|s|^2$):
\begin{equation}
\label{Haus}
a(t)=\frac {\sqrt{2/\tau_e}s}{j(\omega-\omega_0)+1/\tau}[e^{j\omega t}-e^{(j\omega_0-1/\tau)t}].
\end{equation}

\begin{figure}
\centerline{\includegraphics[width=8.3cm]{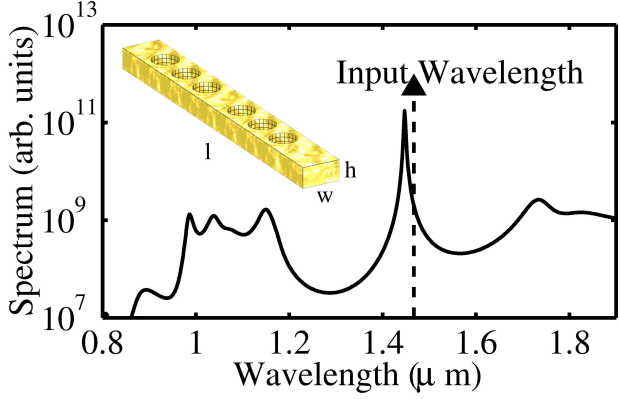}}
\caption{Spectrum of the electric field (y component) probed inside the cavity. The arrow indicates the out-of-resonance input wavelength. The inset is a sketch of the device geometry.}  
\label{fig1}
\end{figure}

\begin{figure}
\centerline{\includegraphics[width=8.3cm]{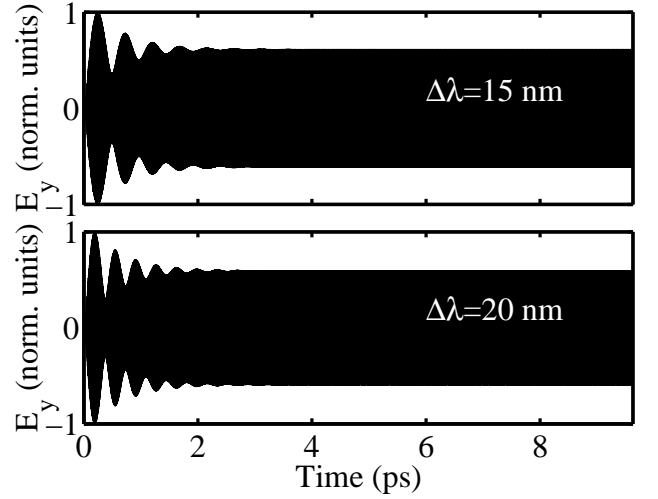}}
\caption{Electric field evolution of the signal inside the cavity when pumped by a step-like excitation at $\lambda=1.462 \mu$m and $\lambda=1.467 \mu$m, corresponding respectively to $\Delta\lambda=15$ nm and $\Delta\lambda=20$ nm.}
\label{fig2}
\end{figure}

Fig. \ref{fig2} displays the time evolution of the y-component of the electric field inside 
the cavity for two different wavelengths. 
The top panel corresponds to an excitation at $\Delta\lambda=\lambda_{input}-\lambda_0=15nm$, the bottom one at $\Delta\lambda=20 nm$. 
As expected from Eq. (\ref{Haus}) the larger the shift from resonance, the higher the frequency of the oscillating transient. 
THz waves result from optical rectification of such transient inside the cavity, 
due to the locally pronounced radiation intensity.
Such generation can be written in term of the quadratically nonlinear polarization $P^{j}=\chi^{jkl}E_{k}E_{l}^*$ 
(using contracted notation),
with $\chi^{jkl}$ the pertinent component of the susceptibility tensor, mixing intra-cavity fields $E_{k}$ and $E_{l}$, 
corresponding to the spatial components $f_{k}({\bf r})$ and $f_{l}({\bf r})$of the defect mode, respectively. 
Naming $\delta=\omega_0-\omega$ the frequency generated via the interaction,
the resulting amplitude of the THz nonlinear polarization is:

\begin{equation}
\label{PTHZ2}
P^{j}_{THz}=\chi^{jkl} f_k f_l^* Q_o \omega_0  P_{input} \frac {\tau_{o}/\tau_e} {(1+2\tau_{o}/\tau_e)^{2}+\delta^{2}\tau_o^{2}}.
\end{equation}

\begin{figure}
\centerline{\includegraphics[width=8.3cm]{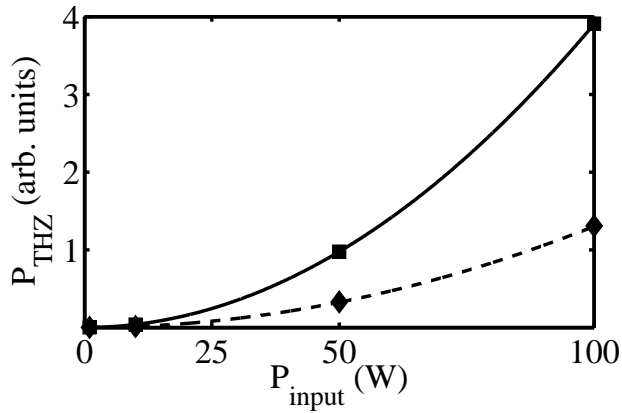}}
\caption{THz generated power at $\lambda=1.462 \mu$m (squares) and $\lambda =1.467 \mu$m (diamonds) versus input power.}
\label{fig3}
\end{figure}

\begin{figure}
\centerline{\includegraphics[width=8.3cm]{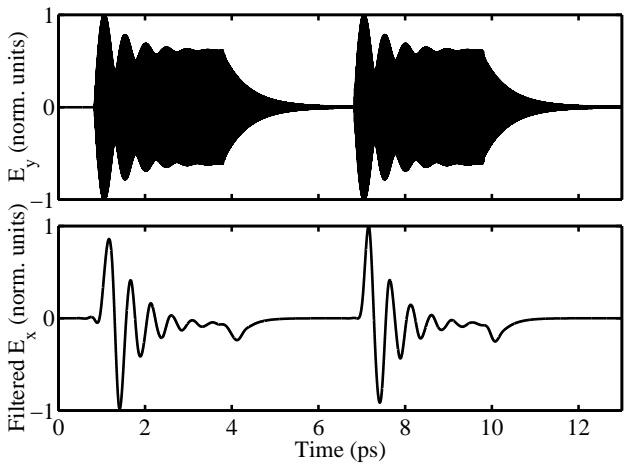}}
\caption{Two mnm pulses at $\lambda=1.462 \mu m$ (top panel) giving rise to two THz pulses after filtering (bottom panel).}
\label{fig4}
\end{figure}

The temporal dynamics is governed by the cavity decay-time, and provides few-cycle pulses at THz. 
Since the latter wavelengths are in the millimeter range, the radiation is unconfined and 
barely ``senses'' the structure. If required, however, it could be trapped e.g. by a hollow pipe waveguide. \cite{Mann}
Fig. \ref{fig3} displays the generated THz power evaluated by FDTD from the power density spectra (PDS) and at two detuning $\delta$.
The curves clearly exhibit a quadratic dependence from the input power at $\omega$, the lines being quadratic fits.\\
While the final conversion efficiency depends on the (eventually) adopted THz guiding geometry, 
the case $\lambda=1.462 \mu$m (corresponding to $\delta=2.1 THz$) is clearly more efficient than for 
$\lambda=1.467 \mu$m ($\delta=2.8 THz$), as expected (see Eq. \ref{PTHZ2}).\\
To demonstrate how this scheme works for a pulsed excitation, in our FDTD code we fed the input waveguide with two ``mnm'' pulses ($m=6$, $n=600$), 
mimicking a Return-to-Zero modulation of the optical pump. 
The top panel of Fig. \ref{fig4} shows the $y$ component of the electric field inside the cavity when pumped 
at $\lambda=1.462 \mu m$.
To extract the THz portion of the electromagnetic field we performed the spectrogram of the TM component, allowing it to follow
the time evolution of the signal at the desired frequency. 
The bottom panel of Fig. \ref{fig4} plots the THz pulses obtained through this procedure.\\
Finally, to quantify the generation efficiency $\eta$, we evaluated the intensity of both optical and THz signals, 
integrating the PDS of TE and TM components around $\omega_0$ and $\delta$, respectively. This yielded a promising
$\eta=2.3 \times 10^{-7}$ for $\delta=2.1 THz$ and $\eta=1.3 \times 10^{-7}$ for $\delta=2.8 THz$, respectively. 
Note that, in this unoptimized 3-hole PC wire, $Q=373$ is far below the values recently obtained in PC cavities, as high as 
$Q=1\times 10^{6}$. \cite {Ryu,Zhang} 

The best performance can be achieved when $\tau_e$ maximizes the last factor in Eq. (\ref{PTHZ2}), hence THz generation becomes linearly dependent on $Q_o$ and the final efficiency grows quadratically to values 
competitive with those recently achieved. \cite{Shi}\\
In conclusion, we numerically investigated THz sources based on optical rectification 
of the transient response of a PC-wire microcavity. 
This optically pumped device, amenable to be incorporated in 
THz circuitry, can be tailored to specific requirements in terms of frequency oscillation and efficiency.


\bibliographystyle{osajnl}

\end{document}